\documentclass[prc,twocolumn,showpacs]{revtex4}

\input{epsf}

\usepackage{graphicx}
\usepackage{dcolumn}
\usepackage{bm}

\usepackage{amsmath}

\begin{document}

\title{
Residual interaction effects on deeply bound pionic  
states in Sn and Pb isotopes}

\author{N. Nose-Togawa,$^1$
H. Nagahiro,$^2$
S. Hirenzaki,$^2$
and K. Kume$^2$}
\affiliation{
$^1$Research Center for Nuclear Physics (RCNP), Ibaraki, Osaka 567-0047, Japan\\
$^2$Department of Physics, Nara Women's University, Nara
630-8506, Japan
}%

\date{\today}

\begin{abstract}
We have studied the residual interaction effects 
theoretically on the deeply bound pionic states in Pb and Sn isotopes. 
We need to evaluate the residual interaction effects carefully in order to deduce 
the nuclear medium effects for pion properties, which are believed to 
provide valuable information on nuclear chiral dynamics. The $s$- and 
$p$-wave $\pi N$ interactions are used for the pion-nucleon residual 
interactions.  
We show that the complex energy shifts are
around [(10-20)+$i$(2-7)]
keV for $1s$ states in Sn, which should be taken into account
in the analyses of 
the high precision data of deeply bound pionic $1s$ states in Sn isotopes.
\end{abstract}

\maketitle
\section{Introduction}

Deeply bound pionic states in heavy nuclei were predicted to be
 quasistable by Friedman and Soff~\cite{Friedman85}, 
 and Toki and Yamazaki~\cite{TokiYama88} independently.  
According to the 
theoretical predictions for the formation reactions ~\cite{Toki91,Hirenzaki91}, 
the deeply bound pionic $2p$ states are 
 observed in $^{207}$Pb nucleus experimentally 
in (d,$^3$He) missing mass spectra in 1996~\cite{Yamazaki96,ItaGil00}. 
After this discovery, precise data of the deeply bound pionic $1s$ and $2p$ states 
in $^{205}$Pb were also observed~\cite{Hirenzaki97,Geissel02}.  
Furthermore, Umemoto $et$ $al$. predicted that the Sn isotopes are ideal 
target nuclei to observe $1s$ pionic states and to deduce the isotope 
shifts of the pionic atoms~\cite{Umemoto00}.  
Recently,
 K. Suzuki $et$ $al$. performed the experiments of the (d,$^3$He) reactions on 
 the Sn targets and 
succeeded to observe deeply bound $1s$ pionic states in Sn isotopes quite 
precisely~\cite{Suzuki04}.  Experimental errors for the 
binding energies of the $1s$ states 
are 
around $\Delta E 
\sim 20 {\rm keV}$.  

From these experiments, we can study the s-wave part of the
pion-nucleus interaction, which is very interesting 
since the $s$-wave strength is expected to provide
information on the pion mass excess   
and pion decay constant $f_\pi$ in the nuclear medium 
through the Tomozawa-Weinberg theorem~\cite{Tomozawa66,Weinberg66}. 
The $f_\pi^2$ is the order parameter of chiral symmetry 
breaking of QCD, and is connected to the quark condensate through the 
Gell-Mann-Oakes-Renner relation~\cite{Gellmann68}. 
Thus, it is very interesting to 
determine the $s$-wave potential parameters from deeply bound pionic 
atoms.  For this purpose, it is required to observe the pionic 
$1s$-states because these states depend predominately on
the $s$-wave potential~\cite{Umemoto00} and,    
furthermore, the $1s$ states in heavy nuclei ($N > Z$) provide key information
on the 
isovector part of the $s$-wave potential.  
As described above, K. Suzuki $et$ $al$. performed the experiment 
and obtained excellent new data of 
the deeply bound pionic $1s$ states in Sn isotopes~\cite{Suzuki04},
which are very suited for the purpose.  

However, since we make use of the single neutron pickup (d,$^3$He) 
reactions, the final pionic states are the 
pion plus neutron-hole state 
$[\pi\otimes n^{-1}]_J$ 
with respect to the target nuclei~\cite{Hirenzaki91,Hirenzaki97,Umemoto00}. 
So far all theoretical calculations and analyses of the data, 
except for Ref.~\cite{Hirenzaki99}, 
postulate that the residual interaction effects are small enough and can be neglected. 
This is actually true for pionic atoms in $^{207}$Pb case since 
experimental errors are
significantly larger 
than the estimated residual interaction effects~\cite{Hirenzaki99}.  
However, in the present 
cases for the $1s$ states in Sn isotopes,  it is not 
obvious whether the effects are negligible or not since the 
experimental errors for Sn cases are comparable 
to the calculated residual interaction effects for
$^{207}{\rm Pb}$~\cite{Hirenzaki99}.  
Thus, it is very important to evaluate the residual-interaction
effects 
for $1s$ states in Sn isotopes to deduce physical quantities related to 
pion behaviors in the nuclear medium from the observed spectra.  
In this report we evaluate the residual-interaction effects on 
pionic states in $^{207}{\rm Pb}$, $^{205}{\rm Pb}$ and Sn isotopes 
by taking into account both $s$-wave and $p$-wave $\pi$-N interactions.  

In section 2, we describe the theoretical formula to evaluate the 
residual interaction effects.
 In section 3, we show the numerical results and 
section 4 is devoted to the conclusions.  

\section{Formulation}

We consider the pionic states whose Hamiltonian is expressed as follows;
\begin{equation}
H =\sum_i \omega _i c^+ _i c_i + \sum_i \varepsilon _i
 a^+_ia_i
+\sum_{ijk\ell}V_{ji,\ell k}c^+_ja^+_ia_k c_\ell\ ,  
\label{eq:Hamiltonian}
\end{equation}
where the $c^+ \ (c)$ and $a^+ \ (a)$ are creation (annihilation) operators
of the pion and the nucleon respectively. The indices characterize
their quantum numbers.
In Eq.~(\ref{eq:Hamiltonian}),
$\omega_i$ is the pion binding energy
, $\varepsilon_i$ the single-nucleon energy and $V_{ji,\ell k}$ 
indicate the matrix elements of the pion-nucleon residual interaction. 

Since we make use of the single-neutron pickup $({\rm d,^3 He})$ reaction,
each final state is the pion plus single neutron-hole state with respect 
to the target nucleus. To calculate the residual-interaction effects 
between pion and the neutron-hole, we introduce a neutron-hole 
creation (annihilation) operator $b^\dag (b)$ 
, which are defined as;
\begin{equation}
a^+_{jm}=(-1)^{j-m}b_{j-m}\ .
\end{equation}
Here, we show the angular momentum
quantum numbers explicitly while the isospin indices are
 abbreviated. The third term in Eq.(1) can be rewritten as
\begin{eqnarray}
\sum_{ijk\ell}V_{ji,\ell k}c^+_ja^+_ia_k c_\ell\ 
\longrightarrow 
   \hat{V} = \sum_{ijk\ell}\overline{V_{jk,\ell i}}c^+_jb^+_kb_ic _\ell\ ,
\end{eqnarray}
where we discarded the core contribution which 
is already included in the second term of Eq.~(\ref{eq:Hamiltonian}) and
$\overline{V_{jk,\ell i}}$ are the interaction matrix elements between 
pion and the nucleon hole which correspond to the Pandya transformation 
of the pion-nucleon interaction. 

The state of pionic atom with a single neutron-hole state can be expressed as
\begin{equation}
|\pi,N_\alpha ; J\rangle =(c^+_\pi\otimes b^+_{N_\alpha})^J|0\rangle\ ,
\end{equation}
where the suffixes $\pi$ and $N_\alpha$ specify the quantum numbers of the
pion and the neutron-hole respectively,
 and $J$ is the total angular
momentum of the pion-nucleus system.
 The matrix elements of the Hamiltonian with respect to these states
are expressed as
\begin{eqnarray}
\langle\pi^{'},N_\beta ;J| H|\pi,N_\alpha ;J
\rangle
&=&(\omega_\pi-\varepsilon_\alpha
)\delta_{\pi,\pi'}\delta_{\alpha,\beta}\nonumber\\
&+&
\langle\pi^{'},N_\beta;J|\hat{V}|\pi,N_\alpha ;J\rangle \ ,
\label{eq:MatEle}
\end{eqnarray}
where $\omega_{\pi}$ is the eigenenergy of the pionic state
specified as $\pi$, and $\varepsilon _{\alpha}$ is the separation 
energy of neutron from the target nucleus.

As the residual interaction,
we consider the pion-nucleon interaction,
\begin{equation}
V = -\frac{2\pi}{m_\pi}[b_0+b_1 \mbox{\boldmath $\tau$}\cdot
{\bf I}+(c_0+c_1 \mbox{\boldmath$\tau$}\cdot {\bf I}){\bf \nabla \cdot
\nabla}] \delta ({\bf r}).
\end{equation}
Here, we have taken into account both the $s$- and the $p$-wave pion-nucleon interaction.
The gradient operators act on the right and the left-hand-side pion
wave functions, respectively.
In Ref.~\cite{Hirenzaki99}, we reported the 
results with the $s$-wave contribution only.
We fix the parameters as $b_0=-0.0283 m_\pi^{-1}$ ,
$b_1= -0.12 m_\pi^{-1}$,
$c_0= 0.223 m_\pi^{-3}$,
$c_1= 0.25 m_\pi^{-3}$,
which are taken from Ref.~\cite{Seki83}.
The pion-nucleon interaction adopted here is consistent with the 
pion-nucleus optical potential used to calculate the pion
wavefunctions.
By folding the pion-nucleon interaction with the nuclear density,
we obtain exactly the same real part of the pion-nucleus optical
potential except for the small corrections coming from the 
transformation of the center of mass coordinates.
As for the imaginary part, we simply assume the pion-nucleon 
residual interaction is real and has no absorptive effects 
since two nucleon degrees of freedom are necessary at least in absorptive
processes.
The effects of the pion absorption by the core nucleus are incorporated 
phenomenologically as the density quadratic term in the imaginary parts of the 
pion-nucleus optical potential as usual.
In this theoretical framework, we do not evaluate the absorptive 
effects due to processes including both nucleon-hole and
nucleon-particle degrees of freedom simultaneously, which we expect to 
be small.

The interaction matrix elements between the pion and the nucleon hole 
are expressed as,
\begin{widetext}
\begin{eqnarray}
&\ &\langle\pi ',N_\beta;J| \hat{V} |\pi,N_\alpha;J\rangle
\nonumber\\
&=&-\frac{1}{2m_\pi}(-1)^{-J+j_\alpha+j_\beta +1/2}
\sqrt{(2j_\alpha +1)(2j_\beta +1)(2\ell_\alpha +1)(2\ell_\beta +1)(2
\ell^{'}_\pi
+1)(2\ell_\pi +1)}\nonumber\\
&\ & \times  \sum_L (-1)^{L} \left\{ \begin{array}{ccc}
          \ell^{'}_\pi  & j_\beta & J\\
          j_\alpha      & \ell_\pi   & L
         \end{array}
\right\}\left\{\begin{array}{ccc}
          \ell_\alpha & j_\alpha   & \frac{1}{2}\\
          j_\beta    & \ell_\beta & L
\end{array} \right\}
(\ell_\beta 0 \ell_\alpha 0 \mid L 0)
(\ell_\pi  0 \ell^{'}_\pi 0 \mid L 0) \nonumber \\
&\ & \times \left[
 (b_0+b_1)\ \int^\infty_0
drr^2R^*_{\ell_\beta}(r)R_{\ell_\alpha}(r)R_{\ell^{'}_\pi}(r)R_{\ell_\pi}(r) \right.
\nonumber\\
&\ &\ \ \ \ \  +(c_0+c_1)\ \int^\infty_0
drr^2R^*_{\ell_\beta}(r)R_{\ell_\alpha}(r)\nonumber\\
&\ &
\left.
 \ \ \ \ \ \times \left\{\left(\frac{{\rm
d}R_{\ell _\pi '}(r)}{{\rm d}r}\right)\left(\frac{{\rm
d}R_{\ell_\pi}(r)}{{\rm d}r}\right)+\frac{\ell_{\pi}(\ell_{\pi}+1)+\ell^{'}_{\pi
}(\ell^{'}_{\pi}+1)-L(L+1)}
{2}\frac{R_{\ell _\pi '} (r)R_{\ell_\pi}(r)}{r^2} \right\}
\right]\ \ .
\end{eqnarray}
\end{widetext}
\noindent
Where $R_{\ell_\pi}(r)$ and $R_{\ell_\alpha}(r)$ are the radial wave
function of the pion and the neutron-hole, respectively.
We consider the pionic orbits of $1s$, $2s$, $2p$, $3s$, $3p$ and
$3d$ states which are obtained
by solving the Klein-Gordon equation numerically.
Because the Klein-Gordon equation includes the complex optical potential which 
makes the Hamiltonian non-Hermite and makes the eigenenergies complex,
and hence we normalize the pionic wave function on the proper orthonormal condition
 according to the prescription in Ref.~\cite{Nose97}.

For the proton and the neutron distributions, we use the two-parameter
Fermi type density distribution as,
\begin{equation}
\rho_{p(n)}=\frac{\rho_0}{1+{\rm exp}[(r-r_{p(n)})/a_{p(n)}]}\ ,
\end{equation}
and assume the same radius parameters of the proton and
the neutron. 
These radius parameters and the proton diffuseness parameter
are taken from the experimental values in Ref.~\cite{Fricke95}.
For the diffuseness parameter of the neutron
 we adopt the values in Ref.~\cite{Trzcin01}.
These density parameters are compiled in Table~\ref{table:density}.

\begin{table}[h]
\begin{center}

\begin{tabular}{crrr} \hline\hline
 Nucleus   & $r_p(=r_n)$[fm] & $a_p$[fm] & $a_n$[fm]\\ \hline
$^{116}{\rm Sn}$ & 5.417 & 0.5234 & 0.5837 \\
$^{120}{\rm Sn}$ & 5.459 & 0.5234 & 0.6014 \\
$^{124}{\rm Sn}$ & 5.491 & 0.5234 & 0.6175 \\
$^{132}{\rm Sn}$ & 5.548 & 0.5234 & 0.6487 \\
$^{206}{\rm Pb}$ & 6.631 & 0.5234 & 0.6389 \\
$^{208}{\rm Pb}$ & 6.647 & 0.5234 & 0.6439 \\ \hline
\end{tabular}
\caption{ Nuclear density parameters used in the
present calculations.
\label{table:density}
}
\end{center}
\end{table}

For the neutron-hole states, we have taken into account the orbits
$p_{1/2}^{-1}$, $f_{5/2}^{-1}$,
$p_{3/2}^{-1}$, $i_{13/2}^{-1}$ for $^{205,207}{\rm Pb}$ and
$d_{3/2}^{-1}$, $s_{1/2}^{-1}$,
$h_{11/2}^{-1}$,$g_{7/2}^{-1}$,
$d_{5/2}^{-1}$ for $^{115,119,123,131} {\rm Sn}$.
These states are calculated using a potential of Woods-Saxon form in
Ref.~\cite{Koura00}. 
The neutron separation energies $\varepsilon_\alpha$ are determined from 
experimental data as far as possible.  
We can disregard the spreading widths of the neutron
hole states which are considerably narrower than the width of
the pionic states and little affect the results here. 
For $^{207}{\rm Pb}$, the  
separation energies are obtained from the isotope table~\cite{Isotope} 
as the excited energies of the levels coupled to the neutron pick-up reactions.  
For open shell nuclei $^{205}{\rm Pb}$ and $^{115,119,123}{\rm Sn}$, we adopt the 
excited states observed in a single neutron pick-up
reactions~\cite{Tickle69,Schneid67} and use the observed  
excitation energies to deduce the neutron separation energies.  
In the case that there exist plural states assigned to the same spin and parity, 
we choose the level which has a larger spectroscopic factor. 
As for $^{131}{\rm Sn}$, we use the separation energies deduced from the 
systematics in Ref.~\cite{Umemoto01} since no data of the neutron pick-up 
reactions are available.  
We diagonalize the matrix elements of the whole Hamiltonian expressed in
(\ref{eq:MatEle}). 
Then, we can calculate the complex energy shifts defined as,
\begin{equation}
\Delta E \equiv E(\pi,N_\alpha;J)-(\omega_\pi-\epsilon_\alpha) ,
\end{equation}
where $ E(\pi,N_\alpha ;J) $ are the corresponding eigenenergies of the pion-nucleus
system.
\section{Numerical Results}

In this section, we show the numerical results of the
residual-interaction effects for the pionic atoms.
As we explained above, we include 6 pionic states and 4(5) neutron
states for Pb(Sn) isotopes in the present calculation to evaluate the
matrix elements. Since the residual-interaction effects are larger  
for deeper bound pionic states, we will show the numerical results for pionic $1s$ 
and $2p$ states.  

In Table~\ref{tab:2}, we show the complex energy shifts of the pionic states on 
$^{207}$Pb.  In order to see the contributions from $\pi$N $p$-wave 
interactions, which are newly included in present work, we show both
results with only $s$-wave interaction, and with $s$- and $p$-wave residual 
interactions.  Here, since we have used more realistic neutron wave functions 
and nuclear density distributions than those used in
Ref.~\cite{Hirenzaki99}, the present results are slightly different from
those  in the previous work.
In Table~\ref{tab:2}, the results only with the $s$-wave residual interaction are  
written in the parentheses
and have the same negative sign for all configurations  which means that
the $s$-wave residual interaction effects makes the bound states deeper  
and the level widths
wider. This fact can be understood intuitively as the result of the  
lack of the repulsive $s$-wave interaction
from the removed one neutron.  The calculated results with both $s$- and  
$p$-wave residual interactions are
also shown in the same Table.  We find that the $p$-wave interaction has  
the opposite effects to the $s$-wave interaction
in general and the complex energy shifts become less attractive and  
absorptive in almost all configurations
except for a few cases.  This tendency also can be understood as the  
result of the missing attractive $p$-wave
interaction from one picked-up neutron.  In this case with $^{207}$Pb  
nuclei, the calculated shifts are reasonably smaller
than the experimental errors and we think we can safely neglect the residual  
interaction effects as concluded in Ref.~\cite{Hirenzaki99}.

We show the calculated results in Table~\ref{tab:3} for pionic atoms in $^{205}$Pb  
with the experimental errors reported in Ref.~\cite{Geissel02}.
In this case, the largest shifts appears for $[ (2p)_\pi \otimes  
f_{5/2}^{-1}]_{3/2}$ configuration and is around half of the
corresponding experimental error for the real part.  However, this  
configuration only has minor contribution to the formation cross  
section~\cite{Hirenzaki97,Geissel02}.  The dominant contribution to the
formation process of pionic $2p$   
state is from $[ (2p)_\pi \otimes p_{3/2}^{-1}] $ configurations  
and
the residual-interaction shifts for this configuration are evaluated to
be around 1/3 or less of the experimental error in real part.
The each level
corresponding to different total angular momentum $J$ has  
different energy shifts and splits by a few keV, which will  
be seen as
a broadening  of the resonance peak since each level overlaps due to their  
large natural widths.
As for the pionic $1s$ state,
the residual-interaction effects are around 1/4 $\sim$ 1/5 of  
experimental error for real part and smaller by about 10 for imaginary
part.
Therefore, we can also conclude that the residual interaction effects
can be neglected safely for pionic atoms in $^{205}$Pb case.

%
%
For Sn isotopes, we have made similar calculation for $^{115}{\rm
Sn},^{119}{\rm Sn}$ and $^{123}{\rm Sn}$.
In these cases together with $^{205}$Pb case,
the target nuclei are not closed and thus the description of the
nuclear structure is much complicated. The purpose of the present 
calculation is, however,
  not to make detailed comparison with the experiment but to estimate the 
size of the
correction coming from the effects of the residual interaction between pion
and the residual nucleus. Then, we simply assumed that the residual nucleus
  $\Psi_r$ consists of a single-hole state with respect to the target nucleus
  $\Psi_i$ as,
\begin{equation}
\Psi_r  = C b_{\alpha}^+ \Psi_i,
 \label{eq:psi}
\end{equation}
and we simply assumed the constant $C=1$ to estimate the largest possible
residual-interaction effects.
The calculated results for the $^{115-123}{\rm Sn}$ are compiled in
Table~\ref{tab:115Sn_etc}. As can be seen, 
  the residual interaction shifts are comparable to the experimental
errors of real part for the pionic $1s$ states, which are most important states
and have dominant contributions for the formation
reaction~\cite{Umemoto00,Suzuki04}. Typically, the real 
  energy shifts are around 15 keV and the imaginary shifts are around
5 keV for pionic $1s$ states.
The residual interaction effects slightly decreases for heavier Sn isotopes,
since the binding energies of $1s$ pionic states are smaller and less
bound for heavier Sn isotopes~\cite{Umemoto00}.
We also show the calculated results for the pionic states in
$^{131}$Sn in Table~\ref{tab:131Sn}, which has the single neutron-hole
configuration with respect to 
the doubly closed-shell structure.
The results are close to those of the 
other Sn
isotopes.
%

In the pionic atom formation spectra of the (d,$^3$He) reactions on Sn
targets, the dominant configuration is $[(1s)_\pi\otimes
s_{1/2}^{-1}]_{1/2}$~\cite{Umemoto00}. Thus, the residual interaction
effects of this configuration should be considered carefully. As we can
see in Table~\ref{tab:115Sn_etc}, the residual interaction effects on
the $[(1s)_\pi\otimes s_{1/2}^{-1}]_{1/2}$ configuration are 
slightly smaller
than experimental error for all isotopes. Hence, we could just manage to
neglect the residual interaction effects again.
 However, the magnitude of the real energy shifts are more or less
comparable 
to those of the experimental error and should be taken into account
seriously in analyses of data with higher precision than Ref.~\cite{Suzuki04}.

\begin{table}[h]
\begin{center}

\begin{tabular}{|c|c|l|c|}
\noalign{\hrule height 0.8pt}
\ & $1s$&\multicolumn{2}{c|}{$2p$}\\
\noalign{\hrule height 0.8pt}
\ &\ &J=1/2& $-7.7-2.1i$\\
$p_{1/2}^{-1}$&$-13.3-2.9i$ & \ &$(-9.4-2.7i)$\\
\cline{3-4}
\ &$(-14.2-3.1i)$& J=3/2 & $-7.7-2.1i$\\
\ &\ &\ &$(-9.3-2.7i)$\\
\hline
\ &\ &J=1/2& $-15.8-4.4i$\\
\ &\ & \ &$(-17.6-5.0i)$\\
\cline{3-4}
$p_{3/2}^{-1}$ &$-12.9-2.9i$&J=3/2&$-0.16+0.2i$\\
\ &$(-13.8-3.1i)$&\ &$(-1.7-0.5i)$\\
\cline{3-4}
\ &\ &J=5/2&$-8.9-2.4i$\\
\ &\ &\ &$(-10.1-3.0i)$\\
\hline
\ &\ &J=3/2&$-13.9-4.6i$\\
\ &\ &\ &$(-15.9-5.2i)$\\
\cline{3-4}
$f_{5/2}^{-1}$&$-13.1-3.5i$&J=5/2&$0.90+0.5i$\\
\ &$(-14.1-3.6i)$&\ &$(-0.8-0.3i)$\\
\cline{3-4}
\ &\ &J=7/2&$-9.5-3.1i$\\
\ &\ &\    &$(-11.4-3.8i)$\\
\hline
\ &\ &J=11/2&$-13.4-7.6i$\\
\ &\ &\ &$(-17.2-7.4i)$\\
\cline{3-4}
$i_{13/2}^{-1}$ &$-14.8-6.3i$&J=13/2&$2.1+1.0i$\\
\ &$(-17.2-5.7i)$&\ &$(-0.2-0.1i)$\\
\cline{3-4}
\ &\ &J=15/2&$-11.3-6.5i$\\
\ &\ &\ &$(-14.9-6.4i)$\\
\hline
Exp. & &\multicolumn{2}{c|}{$\pm 20{\rm (stat.)}\pm120{\rm (sys.)}$} \\
error & &\multicolumn{2}{c|}{$\pm 30i{\rm (stat.)}\pm30i{\rm (sys.)}$} \\
\noalign{\hrule height 0.8pt}
\end{tabular}
\caption{Calculated complex energy shifts due to the residual
interaction in $^{207}{\rm Pb}$.
The results are shown in units of keV for $[(1s)_{\pi}\otimes
j_n^{-1}]_J $ and $[(2p)_{\pi}\otimes j_n^{-1}]_J $ including the
$s$-wave and the $p$-wave parts of pion neutron-hole residual
 interaction. The values in the parentheses 
are the results obtained only with the $s$-wave residual interaction.
Experimental errors are taken from Ref.~\cite{ItaGil00}.
\label{tab:2}
} 
\end{center}
\end{table}

\begin{table}[h]
\begin{center}

\begin{tabular}{|c|c|l|r|}
\noalign{\hrule height 0.8pt}
\ & $1s$&\multicolumn{2}{c|}{$2p$}\\
\noalign{\hrule height 0.8pt}
$p_{1/2}^{-1}$&$-13.6-3.1i$&J=1/2& $-8.3-2.5i$\\
\cline{3-4}
\ &\ & J=3/2 & $0.4+0.2i$\\
\hline
\ &\ &J=1/2& $-15.7-4.4i$\\
\cline{3-4}
$p_{3/2}^{-1}$ &$-13.2-3.1i$&J=3/2&$-0.1+0.3i$\\
\cline{3-4}
\ &\ &J=5/2&$-9.1-2.5i$\\
\hline
\ &\ &J=3/2&$-22.6-7.3i$\\
\cline{3-4}
$f_{5/2}^{-1}$&$-13.5-3.7i$&J=5/2&$0.9+0.6i$\\
\cline{3-4}
\ &\ &J=7/2&$-9.8-3.3i$\\
\hline
\ &\ &J=11/2&$-13.9-8.0i$\\
\cline{3-4}
$i_{13/2}^{-1}$ &$-15.4-6.6i$&J=13/2&$2.2+1.1i$\\
\cline{3-4}
\ &\ &J=15/2&$-11.7-6.8i$\\
\noalign{\hrule height 0.8pt}
Exp. error&$\pm 61\begin{array}{c}
                        +86i\\
                        -77i
\end{array}$ & \multicolumn{2}{c|}{
$\pm 45 \begin{array}{c}
	+30i\\
	-31i
\end{array}$} \\
\noalign{\hrule height 0.8pt}
\end{tabular}

\caption{Calculated complex energy shifts due to the residual
interaction in $^{205}{\rm Pb}$.
The results are shown in units of keV for $[(1s)_{\pi}\otimes
j_n^{-1}]_J $ and $[(2p)_{\pi}\otimes j_n^{-1}]_J $ including the
$s$-wave and the $p$-wave parts of pion neutron-hole residual
 interaction.
Experimental errors are taken from Ref.~\cite{Geissel02}.
\label{tab:3}
}

\end{center}
\end{table}

\section{Conclusion}

In summary, we have evaluated the complex energy shifts of the 
deeply bound pionic states
due to the residual interaction
in Pb and Sn isotopes. We have shown the numerical
results
which include both $s$-wave and $p$-wave $\pi N$ residual-interaction
effects. 
For the open-shell nuclei,
we have assumed  
one-neutron-hole configuration as described in
Eq.~(\ref{eq:psi}).   
The present
results show that the sizes of the residual-interaction effect are 
slightly smaller than the 
experimental errors
in Ref.~\cite{Suzuki04}
for $1s$ pionic states in Sn
isotopes. 
Hence, we could conclude that we can neglect the residual interaction
effects in the analyses of data in Ref.~\cite{Suzuki04} for deeply bound
pionic $1s$ states in Sn as in the cases of Pb. However, the magnitude
of the residual interaction effects are more or less comparable to the experimental
errors in the latest data and the effects should be taken into account
seriously in analyses of data with higher accuracy than
Ref.~\cite{Suzuki04}. We think that it is essentially important 
to study deeply bound pionic atoms in future to deduce
quantitative information on nuclear chiral dynamics.


\section{Acknowledgments}

We would like to thank Dr.~Hirono Fukazawa for useful discussions on the
residual interaction effects on deeply bound pionic states. This work is
partly supported by the Grant-in-Aid for Scientific Research (c) 16540254.

\begin{widetext}

\begin{table}[h]
\begin{center}

\small
\begin{tabular}{|c|c|l|r|c|l|r|c|l|r|}
\noalign{\hrule height 0.8pt}
\ & \multicolumn{3}{c|}{\large $^{115}{\rm
Sn}$}&\multicolumn{3}{c|}{\large $^{119}{\rm Sn}$} &
\multicolumn{3}{c|}{\large $^{123}{\rm Sn}$}\\
\cline{2-10}
\ & $1s$&\multicolumn{2}{c|}{$2p$}&$1s$&\multicolumn{2}{c|}{$2p$}&$1s$&\multicolumn{2}{c|}{$2p$}\\
\noalign{\hrule height 0.8pt}
$s_{1/2}^{-1}$&$-15.4$&\footnotesize{J=1/2}& $-4.0-1.1i$&
$-13.5$ & \footnotesize{J=1/2} & $ -5.2-2.0i$ & 
$-12.3$&\footnotesize{J=1/2}& $-3.2-0.7i$\\
\cline{3-4}\cline{6-7}\cline{9-10}
\large{\quad}&\ $\ \ -4.2i$&\footnotesize{J=3/2}& $-4.0-1.1i$&$\ \ -3.3i$&\footnotesize{J=3/2}& $-3.8-1.1i$ &$\ \
-2.4i$&\footnotesize{J=3/2}&
$-3.5-0.8i$\\
\hline
$d_{3/2}^{-1}$ &$-15.9$&\footnotesize{J=1/2}& $-9.1-3.1i$ &$-14.3$&\footnotesize{J=1/2}&$-7.0-1.6i$ &$-12.8$
&\footnotesize{J=1/2}&$-8.1-2.5i$\\
\cline{3-4}\cline{6-7}\cline{9-10}
\large{\quad}&$\ \ -4.8i$&\footnotesize{J=3/2}&$0.3+0.3i$& $\ \ -3.7i$
&\footnotesize{J=3/2}&$0.4+0.3i$&$\ \ -2.9i$&\footnotesize{J=3/2}&$0.2+0.1i$\\
\cline{3-4}\cline{6-7}\cline{9-10}
\large{\quad}&\ &\footnotesize{J=5/2}&$-5.2-1.8i$& \ &\footnotesize{J=5/2}&$-4.6-1.4i$&\ &\footnotesize{J=5/2}&$-4.3-1.2i$\\
\hline
$g_{7/2}^{-1}$&$-15.4$&\footnotesize{J=5/2}&$-6.0-3.8i$&$-13.0$&\footnotesize{J=5/2}&$-5.5-3.3i$&$-11.1$&\footnotesize{J=5/2}&$-4.9-2.8i$\\
\cline{3-4}\cline{6-7}\cline{9-10}
\large{\quad}&$\ \ -7.3i$&\footnotesize{J=7/2}&$1.5+0.8i$&$\ \
-5.8i$&\footnotesize{J=7/2}&$1.3+0.7i$&$\ \ -4.6i$&\footnotesize{J=7/2}&$1.2+0.6i$\\
\cline{3-4}\cline{6-7}\cline{9-10}
\large{\quad}&\ &\footnotesize{J=9/2}&$-4.4-2.9i$&\ &\footnotesize{J=9/2}&$-3.9-2.4i$&\  &\footnotesize{J=9/2}&$-3.5-2.0i$\\ 
\hline
$h_{11/2}^{-1}$&$-18.3$&\footnotesize{J=9/2}&$-7.7-4.0i$&$-16.0$&\footnotesize{J=9/2}&$-6.9-3.5i$&$-14.1$&\footnotesize{J=9/2}&$-6.3-3.0i$\\
\cline{3-4}\cline{6-7}\cline{9-10}
\large{\quad}&$\ \ -7.2i$&\footnotesize{J=11/2}&$1.7+0.8i$&$\ \ -6.0i$&\footnotesize{J=11/2}&$1.5+0.7i$&$\ \ -5.1i$&\footnotesize{J=11/2}&$1.4+0.6i$\\
\cline{3-4}\cline{6-7}\cline{9-10}
\large{\quad}&\ &\footnotesize{J=13/2}&$-6.2-3.3i$&\ &\footnotesize{J=13/2}&$-5.6-2.8i$&\ &\footnotesize{J=13/2}&$-5.1-2.5i$\\
\hline
$d_{5/2}^{-1}$&$-15.1$&\footnotesize{J=3/2}&$-7.6-2.6i$&$-13.6$&\footnotesize{J=3/2}&$-7.1-2.2i$&$-12.2$&\footnotesize{J=3/2}&$-6.5-1.9i$\\
\cline{3-4}\cline{6-7}\cline{9-10}
\large{\quad}&\ $\ \ -4.8i$&\footnotesize{J=5/2}&$1.0+0.6i$&$\ \
-3.7i$&\footnotesize{J=5/2}&$0.9+0.4i$&$\ \ -2.8i$&\footnotesize{J=5/2}&$0.8+0.3i$\\
\cline{3-4}\cline{6-7}\cline{9-10}
\large{\quad}&\ &\footnotesize{J=7/2}&$-5.0-1.7i$&\ &\footnotesize{J=7/2}&$-4.6-1.4i$&\ &\footnotesize{J=7/2}&$-4.3-1.2i$\\
\noalign{\hrule height 0.8pt}
$\begin{array}{c}\rm {Exp.}\\{\rm
error}\end{array}$&$\begin{array}{c}\pm 24\\ \ \ \pm
44i\end{array}$
&\multicolumn{2}{c|}{---------}&$\begin{array}{c}\pm 18\\ \ \ \pm 40i\end{array}$&\multicolumn{2}{c|}{---------}&
$\begin{array}{c}\pm 18 \\ \ \ \pm 36i\end{array}$&\multicolumn{2}{c|}{---------}\\
\noalign{\hrule height 0.8pt}
\end{tabular}
\normalsize
\caption{Calculated complex energy shifts due to the residual
interaction in $^{115,119,123}{\rm Sn}$.
The results are shown in units of keV for $[(1s)_{\pi}\otimes
j_n^{-1}]_J $ and $[(2p)_{\pi}\otimes j_n^{-1}]_J $ including the
$s$-wave and the $p$-wave parts of pion neutron-hole residual
 interaction.
Experimental errors are taken from Ref.~\cite{Suzuki04}.
\label{tab:115Sn_etc}
}
\end{center}
\end{table}

\end{widetext}

\begin{table}[h]
\begin{center}

\begin{tabular}{|c|c|l|r|}
\noalign{\hrule height 0.8pt}
\ & $1s$&\multicolumn{2}{c|}{$2p$}\\
\noalign{\hrule height 0.8pt}
$s_{1/2}^{-1}$&$-10.5-1.3i$&J=1/2& $-3.2-0.6i$\\
\cline{3-4}
\ &\ &J=3/2&$-3.3-0.6i$\\
\hline
\ &\ &J=1/2& $-7.1-2.0i$\\
\cline{3-4}
$d_{3/2}^{-1}$&$-10.4-2.1i$&J=3/2&$0.2+0.0i$\\
\cline{3-4}
\ &\ &J=5/2&$-3.8-1.1i$\\
\hline
\ &\ &J=5/2&$-3.0-1.2i$\\
\cline{3-4}
$g_{7/2}^{-1}$&$-6.5-1.6i$&J=7/2&$\ 0.9+0.4i$\\
\cline{3-4}
\ &\ &J=9/2&$-2.1-0.8i$\\
\hline
\ &\ &J=9/2&$-4.6-1.8i$\\
\cline{3-4}
$h_{11/2}^{-1}$ &$-9.6-2.6i$&J=11/2&$1.1+0.4i$\\
\cline{3-4}
\ &\ &J=13/2&$-3.7-1.4i$\\
\hline
\ &\ &J=3/2&$-5.8-1.5i$\\
\cline{3-4}
$d_{5/2}^{-1}$&$-9.9-1.9i$&J=5/2&$0.6+0.2i$\\
\cline{3-4}
\ &\ &J=7/2&$-3.9-1.1i$\\
\noalign{\hrule height 0.8pt}
\end{tabular}
\caption{Calculated complex energy shifts due to the residual
interaction in $^{131}{\rm Sn}$.
The results are shown in units of keV for $[(1s)_{\pi}\otimes
j_n^{-1}]_J $ and $[(2p)_{\pi}\otimes j_n^{-1}]_J $
including the $s$-wave and the $p$-wave parts of pion neutron-hole residual
 interaction. 
\label{tab:131Sn}
}
\end{center}
\end{table}

\end{document}